\documentclass[a4paper,11pt]{article}
\usepackage{jheppub}
\usepackage[T1]{fontenc}
\usepackage{amsfonts}
\usepackage{amsmath}
\usepackage{epsf}
\usepackage{amssymb}
\usepackage{latexsym}
\usepackage{graphicx,epsfig}
\usepackage{subfigure}
\def\Q{\mathbb{Q}}
\def\M{\mathcal{M}}
\def\C{\mathcal{C}}
\newcommand{\dd}{\mathrm{d}}
\newcommand{\PD}[2]{\ensuremath{\frac{\partial #1}{\partial #2}}}

\title{New Black hole Solutions in $f(\Q)$ Gravity}
\author{A. Dehyadegari and A. Sheykhi}

\affiliation{Department of Physics, College of Science, Shiraz
University, Shiraz 71454, Iran \\ Biruni Observatory, College of
Science, Shiraz University, Shiraz 71454, Iran}

\emailAdd{amin.dehyadegari@hafez.shirazu.ac.ir}
\emailAdd{asheykhi@shirazu.ac.ir}

\abstract{We investigate static and spherically symmetric vacuum
solutions in the symmetric teleparallel $f(\Q)$ modified theory of
gravity. Starting from a recently proposed classification of
affine connections compatible with both the symmetries of
spacetime and the constraints of symmetric teleparallel geometry,
we develop a systematic approach to solve the full field
equations. We first identify two distinct classes of connections
that satisfy the off-diagonal metric field equations and the
connection constraints. For an arbitrary  $f(\Q)$ function when
the non-metricity scalar $\Q$ vanishes, we recover exact
analytical solutions equivalent to those of general relativity,
including the Schwarzschild and Schwarzschild (anti)de-Sitter
metrics. We then extend our analysis beyond general relativity by
considering the quadratic model $f(\Q)=\Q+\alpha~\Q^2$ with a
small parameter $\alpha$. Using a perturbative approach, we derive
asymptotically flat, analytical solutions up to second order in
$\alpha$. These solutions exhibit corrections to the standard
Schwarzschild metric, characterized by new integration constants
that can be interpreted as connection hair. We explore the
asymptotic behavior of these solutions and disclose that the
horizon radius receives corrections that can be expressed
compactly using the Lambert $\mathcal{W}$ function. Our results
provide new, non-trivial vacuum solutions within $f(\Q)$ gravity
and highlight the rich structure introduced by the non-metricity
connection.}
\begin{document}
    \maketitle

 \section{Introduction} \label{sec:Intro}\setcounter{equation}{0}

Generic metric-affine geometry provides a unique framework with the most general form of the metric and the connection. The intrinsic content of such a geometry is encoded in three independent tensorial properties: curvature, torsion, and non-metricity \cite{Eisenhart:1927,Hehl:1995,Blagojevic:2001,Jimenez:2018,Heisenberg:2024}. These fundamental properties play a prominent role in modeling gravity in various reduced geometries. A particularly notable example is the Riemannian geometry, in which general relativity and its extensions are formulated through the curvature tensor. In this setting, the unique torsionless and metric-compatible connection is the Levi-Civita connection. Alternatively, by taking into account the vanishing of curvature and non-metricity, general relativity can emerge solely from the torsion tensor in what is known as the teleparallel equivalent of general relativity (TEGR) \cite{Heisenberg:2024,Hayashi:1979,Maluf:2013}. Extensive studies on TEGR and its various extensions can be found in Refs. \cite{Hohmann:2019,Bahamonde:2020vpb,Hohmann:2019nat,DeBenedictis:2016aze,Ruggiero:2015oka,Ferraro:2018,Li:2011,Blagojevic:2020,Calza:2024}.

Another possible approach to study gravity is to consider the sector of metric-affine geometry in which curvature and torsion are postulated to vanish, while non-metricity is allowed to remain non-zero. In this perspective, general relativity is reformulated in terms of the non-metricity tensor, whose Lagrangian density is given by the non-metricity scalar $\Q$. This theory is referred to as the symmetric teleparallel equivalent of general relativity (STEGR) \cite{Adak:2006,Adak:2013,BeltranJimenez:2017,BeltranJimenez:2019,DAmbrosio:2020,Dambrosio:2020b}. Within the symmetric teleparallel framework, it is possible to construct a new class of extensions of general relativity. Among these extensions, $f(\Q)$ modified gravity has attracted considerable attention, in which the gravitational Lagrangian is a function of the non-metricity scalar $\Q$. $f(\Q)$ theory has been widely investigated in gravitational physics, including cosmology \cite{Dialektopoulos:2019,BeltranJimenez:2020,Barros:2020,Zhao:2022,Heisenberg:2018vsk,Anagnostopoulos:2021,Lymperis:2022,Narawade:2023,Narawade:2024}, astrophysics \cite{Maurya:2024b,Maurya:2024,Gogoi:2023,Wang:2025,Heisenberg:2025}, as well as black hole \cite{Lin:2021,DAmbrosio:2022,Calza:2023,Dimakis:2024} and wormhole solutions \cite{Hassan:2021,Banerjee:2021,Mustafa:2022}.

The study of static and spherically symmetric spacetime configurations within the non-metricity formulation of gravity was carried out in Refs. \cite{Zhao:2022,Lin:2021}, where a connection compatible with these symmetries was first presented. Recently, the authors of Ref. \cite{DAmbrosio:2022} provided a systematic investigation of the general static and spherically symmetric solutions for the metric and connection in $f(\Q)$ gravity. In particular, they derived the set of constraint equations ensuring that the connection is stationary, spherically symmetric, torsionless, and flat, as required by symmetric teleparallel geometry. It was further demonstrated that the connections used in \cite{Zhao:2022,Lin:2021} satisfy these constraint equations and reproduce the metric solutions of general relativity for arbitrary $f$. Then, by deforming this connection, metric solutions beyond general relativity were obtained in $f(\Q)=\Q+\alpha~ \Q^2$ gravity, assuming $\alpha$ to be small. These solutions can endow black holes with an additional connection hair \cite{DAmbrosio:2022}.

In this paper, we study exact static and spherically symmetric vacuum solutions of $f(\Q)$ modified gravity in four-dimensional spacetime. Building on the insights of Ref. \cite{DAmbrosio:2022}, we adopt a slightly modified strategy to solve the constraint equations together with the field equations. We first solve the equations consistently and recover exact analytical solutions equivalent to those of general relativity within symmetric teleparallel gravity. We find four distinct sets of solutions; for one of them, the connection components reduce to those obtained in \cite{Zhao:2022,Lin:2021,DAmbrosio:2022} under a specific condition. Then, adopting the ansatz $f(\Q)=\Q+\alpha~ \Q^2$ with small $\alpha$, we deform the general relativistic branch and construct asymptotically flat solutions beyond general relativity in detail. For each case, the correction to the standard Schwarzschild horizon radius is examined.

The rest of the paper is organized as follows. In the next section (\ref{sec:SymmetricTeleparallelism}), we introduce the basic definitions and notations of symmetric teleparallel $f(\Q)$ gravity. We also present the field equations governing the metric and connection.  In Sec. \ref{sec:SymRed}, we briefly review the general static and spherically symmetric metric and connection ansatz and analyze the corresponding field equations. We then adopt a modified solution strategy compared to that of Ref. \cite{DAmbrosio:2022}. In Section \ref{sec:Solutions}, we first obtain the vacuum solutions equivalent to those of general relativity. We then include a higher-order contribution in the Lagrangian and derive approximate solutions beyond general relativity. We summarize and discuss our results in Sect. \ref{sec:Conclusion}. In the Appendix, we provide a brief introduction to the Lambert function.

\section{Symmetric Teleparallel $f(\Q)$ Gravity}\label{sec:SymmetricTeleparallelism}\setcounter{equation}{0}

To begin with, let us recall the basic framework of metric-affine geometry. The spacetime manifold $\mathcal{M}$ is equipped with a Lorentzian metric $g_{\mu\nu}$ of signature $(-+++)$ and an affine connection $\Gamma^{\alpha}{}_{\mu\nu}$ that is taken to be independent of the metric. In general, the connection does not need to be symmetric with respect to the indices $\mu$ and $\nu$. The covariant derivatives of a vector $V^\alpha$ and a covector $V_\alpha$ are then defined as
\begin{equation}
    \nabla_\mu V^\alpha = \partial_\mu V^\alpha + \Gamma^\alpha_{\ \mu\lambda} V^\lambda,\qquad
    \nabla_\mu V_\alpha = \partial_\mu V_\alpha - \Gamma^\lambda_{\ \mu\alpha} V_\lambda.
\end{equation}
These relations extend directly to arbitrary tensor fields~\cite{Hehl:1995,Jimenez:2018}.
From the affine connection, three independent tensors can be defined: the curvature, the torsion, and the non-metricity. In particular, the curvature tensor is
\begin{equation}\label{eq:curvature}
    R^\alpha_{\ \beta\mu\nu} = 2 \partial_{[\mu}\Gamma^\alpha_{\ \nu]\beta} + 2 \Gamma^\alpha_{[\mu|\lambda|}\Gamma^\lambda_{\ \nu]\beta}.
\end{equation}
It measures how parallel transport around a loop changes the orientation of a vector.
The torsion tensor is
\begin{equation}\label{eq:torsion}
    T^\alpha_{\ \mu\nu} = 2\Gamma^\alpha_{\ [\mu\nu]},
\end{equation}
which is the antisymmetric part of the connection and reflects the failure of parallelogram closure.
Finally, the non-metricity tensor is
\begin{align}
    Q_{\alpha\mu\nu} &= \nabla_\alpha g_{\mu\nu}  \notag \\
    &= \partial_\alpha g_{\mu\nu} - 2\Gamma^{\lambda}{}_{\alpha(\mu}g_{\nu)\lambda},
\end{align}
where it shows that even the length of a vector is altered when it is parallel transported. It is worth noting that, unlike curvature and torsion which are determined only by the connection, the non-metricity tensor also involves the metric.

In the symmetric teleparallel scenario, the affine connection is restricted by two geometric postulates: the curvature and torsion tensors vanish,
\begin{equation}\label{eq:STpostulates}
    R^\alpha_{\ \beta\mu\nu} = 0 \quad\text{and}\quad T^\alpha_{\ \mu\nu} = 0.
\end{equation}
With these conditions, the only geometric quantity that can play a nontrivial role is the non-metricity tensor. Hence, in an alternative geometry approach, gravity can be described entirely in terms of non-metricity rather than the spacetime curvature of General Relativity. In particular, the usual Einstein-Hilbert Lagrangian is given in terms of quadratic combinations of the non-metricity tensor as a non-metricity scalar
\begin{equation}\label{eq:GenNonMetricityScalar}
    \Q= -\frac14\, Q_{\alpha\beta\gamma}Q^{\alpha\beta\gamma} + \frac12\, Q_{\alpha\beta\gamma}Q^{\beta\alpha\gamma}     + \frac14\, Q_\alpha Q^{\alpha} - \frac12 \, Q_\alpha \bar{Q}^\alpha,
\end{equation}
where $Q_\alpha= Q_{\alpha\nu}{}^{\nu}$ and $\bar{Q}_\alpha= Q^\nu{}_{\nu\alpha}$. Indeed, by a direct calculation, it is easy to show that the Ricci scalar of the Levi-Civita connection is related to the non-metricity scalar as follows
\begin{equation}\label{eq:GRequivalentidentity}
    \Q = \mathcal{D}_\mu(Q^\mu-\bar Q^\mu)+\mathcal{R},
\end{equation}
where $\mathcal{D}_\mu$ represents the covariant derivative with respect to the Levi-Civita connection. This implies that the action constructed from the non-metricity scalar, given by
\begin{equation}
    \mathcal S[g, \Gamma]=\frac12 \int_\M \dd^4 x\,\sqrt{-g}\,\Q,
\end{equation}
is equivalent to the Einstein-Hilbert action of general relativity. Therefore, the symmetric teleparallel formalism provides a different geometrical approach to general relativity, allowing for various extensions such as $f(\Q)$ gravity. It is again remarkable that the connection in the above relation is not the Levi-Civita one.

Now, we turn our attention to the nonlinear extension of the non-metricity scalar, namely $f(\Q)$ gravity. This construction is a generalization of General Relativity within symmetric teleparallel gravity. The action of $f(\Q)$ theory in four dimensional spacetime is given by
\begin{equation}\label{eq:CovAction}
    \mathcal S[g, \Gamma] =\frac12 \int_\M \dd^4 x \sqrt{-g}\,f(\Q)  ,
\end{equation}
where $f$ is an arbitrary function satisfying the condition $f'(\Q) = \dd f(\Q) /\dd \Q \neq 0$. This condition ensures that the theory goes beyond general relativity. Taking into account the independence of the metric $g_{\mu\nu}$ and the affine connection $\Gamma^{\alpha}{}_{\mu\nu}$, the variation of the above action with respect to them yields the following field equations, respectively
\begin{align}\label{eq:FieldEquations}
    \M_{\mu\nu} = \frac{2}{\sqrt{-g}}\nabla_\alpha\left[\sqrt{-g}P^\alpha{}_{\mu\nu} f'(\Q)\right] + f'(\Q) q_{\mu\nu} -\frac12 f(\Q) g_{\mu\nu} &= 0\notag\\
    \C_\alpha = \nabla_\mu\nabla_\nu\left(\sqrt{-g}\,f'(\Q) P^{\mu\nu}{}_{\alpha}\right) &= 0,
\end{align}
where $P^\alpha{}_{\mu\nu}$ denotes the non-metricity conjugate and $q_{\mu\nu}$ is the symmetric tensor, defined as
\begin{align}
    P^\alpha{}_{\mu\nu} &= \frac12 \PD{\Q}{Q_\alpha{}^{\mu\nu}} \notag\\
    &= -\frac14 Q^\alpha{}_{\mu\nu} + \frac12 Q_{(\mu}{}^{\alpha}{}_{\nu)} +\frac14 g_{\mu\nu}Q^\alpha -\frac14 \left(g_{\mu\nu} \bar{Q}^\alpha + \delta^\alpha{}_{(\mu} Q_{\nu)}\right)\notag\\
    q_{\mu\nu} &= \PD{\Q}{g^{\mu\nu}} \notag\\
    &= P_{(\mu|\alpha\beta}Q_{\nu)}{}^{\mu\nu} - 2P^{\alpha\beta}{}_{(\nu} Q_{\alpha\beta|\mu)}.
\end{align}
It is worth noting that the metric and connection field equations are related through the identity $\mathcal{D}_\mu \M^\mu{}_{\nu}+\C_\nu=0$ \cite{Heisenberg:2024}.

In the present work, we study the above field equations of $f(\Q)$ gravity for a stationary and spherically symmetric spacetime. Hence, in the next section, we give a brief review on the general form of the metric and affine connection which respect stationarity and spherical symmetry, and are compatible with the symmetric teleparallel gravity postulates.

\section{Symmetry reduction}\label{sec:SymRed}\setcounter{equation}{0}

To obtain analytical solutions in the nonlinear extension $f(\Q)$, we begin by reviewing the most general stationary and spherically symmetric form of the metric and the affine connection. Then, we examine the corresponding components of the field equations for both the metric and the connection. Here, based on the results of Ref. \cite{DAmbrosio:2022}, we provide an alternative approach to solving the field equations in $f(\Q)$ gravity.

\subsection{Metric and connection}\label{subsec:SymRedMC}

To investigate stationary spherically symmetric solutions of the field equations in $f(\Q)$ gravity, one has to first specify the general form of both the metric and the affine connection based on symmetry requirements. In particular, the affine connection must also satisfy the conditions of symmetric teleparallel geometry given in ~\eqref{eq:STpostulates}. Comprehensive details on the derivation of the metric and the connection can be found in Ref. \cite{DAmbrosio:2022}. Here, we only present the main results.

In four dimensions, the general static and spherically symmetric metric of spacetime is given by
\begin{equation}\label{eq:Metric}
    \dd s^{2}=-g_{tt}(r) \dd t^{2}+g_{rr}(r)
    \dd r^{2}+r^{2}\left( \dd \theta ^{2}+\sin ^{2}\theta \dd \phi ^{2}\right),
\end{equation}
where $g_{tt}(r)$ and $g_{rr}(r)$ are components of the metric to be determined by the field equations. In addition to the metric, the affine connection must also be static and spherically symmetric, while fulfilling the constraints of symmetric teleparallel gravity. As discussed in detail in Ref. \cite{DAmbrosio:2022}, two distinct classes of affine connections arise under these conditions. These two cases are summarized in Tables~\ref{tab:SolutionSet1} (solution set 1) and \ref{tab:SolutionSet2} (solution set 2). It is worthwhile to mention that solution set 1 can be obtained from solution set 2 through an appropriate double scaling limit.
\begin{table}[h!]
    \centering
    \begin{tabular}{|p{0.16\textwidth}|p{0.25\textwidth} p{0.16\textwidth} p{0.18\textwidth}|}
        \hline
        & & & \\[-1.5ex]
        \textbf{Independent components} & \multicolumn{3}{l|}{ $\Gamma^{t}{}_{rr}(r), \Gamma^{r}{}_{rr}(r), \Gamma^{\phi}{}_{r\phi}(r)$.}
        \\[1ex] \hline
        & & & \\[-1.5ex]
        &  $\Gamma^{t}{}_{tt} = c$ & $\Gamma^{t}{}_{tr} = \Gamma^{\phi}{}_{r\phi}$ & $\Gamma^{t}{}_{\theta\theta} = -1/c$\\[1ex]
        \textbf{Non-zero components} & $\Gamma^{t}{}_{\phi\phi} = -\sin^2\theta/c$ & $\Gamma^{\theta}{}_{t\theta} = c$ & $\Gamma^{\theta}{}_{r\theta} = \Gamma^{\phi}{}_{r\phi}$ \\[1ex]
        & $\Gamma^{\theta}{}_{\phi\phi} = -\cos\theta\,\sin\theta$ & $\Gamma^{\phi}{}_{t\phi} = c$ & $\Gamma^{\phi}{}_{\theta\phi} = \cot\theta$ \\[1ex]
        & \multicolumn{3}{l|}{Real constant parameter $c\neq 0$.}
        \\[1ex] \hline
        & & & \\[-1.5ex]
        \textbf{Differential relation} & \multicolumn{3}{l|}{$\partial_r \Gamma^{\phi}{}_{r\phi} = c\,\Gamma^t_{rr} + \Gamma^{\phi}{}_{r\phi}(\Gamma^{r}{}_{rr}-\Gamma^{\phi}{}_{r\phi})$.}
        \\[1ex] \hline
    \end{tabular}
    \caption{Independent and representative non-zero components of the affine connection, together with the differential relation, characterizing solution set 1 \cite{DAmbrosio:2022}. These results correspond to the static and spherically symmetric case in four dimensions, consistent with the postulates of symmetric teleparallel gravity.}
    \label{tab:SolutionSet1}
\end{table}

\begin{table}[h!]
    \centering
    \begin{tabular}{|p{0.16\textwidth}|p{0.28\textwidth} p{0.3\textwidth} p{0.22\textwidth}|}
        \hline
        & & & \\[-1.5ex]
        \textbf{Independent components} & \multicolumn{3}{l|}{ $\Gamma^{t}{}_{rr}(r), \Gamma^{t}{}_{\theta\theta}(r),\Gamma^{r}{}_{rr}(r), \Gamma^{r}{}_{\theta\theta}(r)$.}
        \\[1ex] \hline
        & & & \\[-1.5ex]
        & $\Gamma^{t}{}_{tt} = k-c+c(k-2c)\Gamma^{t}{}_{\theta\theta}$ & $\Gamma^{t}{}_{tr} =- \frac{(k-2c)\Gamma^{t}{}_{\theta\theta}(1+c\,\Gamma^{t}{}_{\theta\theta})}{\Gamma^{r}{}_{\theta\theta}}$ & $\Gamma^{t}{}_{\phi\phi} = \sin^2\theta\, \Gamma^{t}{}_{\theta\theta}$ \\[1ex]
        \textbf{Non-zero components}
        & $\Gamma^{r}{}_{tt} = c(k-2c)\Gamma^{r}{}_{\theta\theta}$ & $\Gamma^{r}{}_{tr} = c-c(k-2c)\Gamma^{t}{}_{\theta\theta}$ & $\Gamma^{r}{}_{\phi\phi} = \sin^2\theta\, \Gamma^{r}{}_{\theta\theta}$ \\[1ex]
        & $\Gamma^{\theta}{}_{t\theta} = c$ & $\Gamma^{\theta}{}_{r\theta} = -(1+c\,\Gamma^{t}{}_{\theta\theta})/\Gamma^{r}{}_{\theta\theta}$ & $\Gamma^{\theta}{}_{\phi\phi} = -\cos\theta\,\sin\theta$ \\[1ex]
        & $\Gamma^{\phi}{}_{t\phi} = c$ & $\Gamma^{\phi}{}_{r\phi} = -(1+c\,\Gamma^{t}{}_{\theta\theta})/\Gamma^{r}{}_{\theta\theta}$ & $\Gamma^{\phi}{}_{\theta\phi} = \cot\theta$ \\ [1ex]
        & \multicolumn{3}{l|}{Arbitrary real constants $c$ and $k$, together with the condition $\Gamma^{r}{}_{\theta\theta}(r)\neq 0$.}
        \\[1ex] \hline
        & & & \\[-1.5ex]
        \textbf{Differential relations} & \multicolumn{3}{l|}{$\partial_r \Gamma^t{}_{\theta\theta} = -\frac{\Gamma^t{}_{\theta\theta}}{\Gamma^r{}_{\theta\theta}}\left[1+\Gamma^t{}_{\theta\theta}\left(3c-k-c(k-2c)\,\Gamma^t{}_{\theta\theta}\right)\right] -\Gamma^t{}_{rr}\Gamma^r{}_{\theta\theta}$,} \\ [1ex]
        & \multicolumn{3}{l|}{ $\partial_r \Gamma^r{}_{\theta\theta} = -1 - c\, \Gamma^t{}_{\theta\theta}\left(2-(k-2c)\,\Gamma^t{}_{\theta\theta}\right)-\Gamma^r{}_{rr}\Gamma^r{}_{\theta\theta}$.}
        \\[1ex] \hline
    \end{tabular}
    \caption{Independent and representative non-zero components of the affine connection, together with the differential relation, characterizing solution set 2 \cite{DAmbrosio:2022}. These results correspond to the static and spherically symmetric case in four dimensions, consistent with the postulates of symmetric teleparallel gravity.}
    \label{tab:SolutionSet2}
\end{table}
In the next section, making use of these solutions, we present the form of the metric and connection field equations. We then discuss in detail the field equation components associated with each solution set. Our aim is to seek solutions that go beyond those of Einstein general relativity.

\subsection{Field equations for the metric and the connection}\label{subsec:SymRedFieldEq}

Having established the metric and the two distinct classes of affine connections compatible with staticity, spherical symmetry, and the postulates of symmetric teleparallel gravity, we now proceed to address the corresponding field equations. Within this framework, the possibility of obtaining solutions that go beyond those of general relativity is examined. Configurations that merely reproduce the known solutions are not considered here.

According to the metric~\eqref{eq:Metric} and solution sets 1 (Table~\ref{tab:SolutionSet1}) and 2 (Table~\ref{tab:SolutionSet2}),
the vacuum field equations for both cases take the following matrix form:
\begin{align}
    &\text{Reduction of metric field equations: }
    &\M_{\mu\nu} &= \begin{pmatrix}
        \M_{tt} & \M_{tr} & 0 & 0\\
        \M_{tr} & \M_{rr} & 0 & 0\\
        0 & 0 & \M_{\theta\theta} & 0\\
        0 & 0 & 0 & \M_{\theta\theta}\,\sin^2\theta
    \end{pmatrix}, \notag\\
    &\text{Reduction of connection field equations: }
    &\mathcal C_{\alpha}& =\begin{pmatrix}
        \mathcal C_t \\
        \mathcal C_r \\
        0 \\
        0
    \end{pmatrix}.
\end{align}
As can be seen, there are four independent components of the metric field equations and two independent components of the connection equations. Following the strategy adopted in Ref. \cite{DAmbrosio:2022}, we start from the off-diagonal metric equation for both solution sets in order to identify new solutions that are distinct from the standard ones of general relativity. In other words, it is possible to determine which set of connection solutions leads to non-trivial results in in $f(\Q)$ gravity.

\subsubsection{Off-diagonal metric equation for solution set 1}\label{ssec:SymRedFieldEq1}

For solution set 1, the off-diagonal part of the metric equations has the simple form
\begin{equation}\label{eq:OffDiagSolset1}
    \M_{tr} = \frac32 c\,\partial_r\Q\, f''(\Q) = 0.
\end{equation}
With regard to $c\neq 0$ in solution set 1 (see Table~\ref{tab:SolutionSet1}), the equation can only be satisfied if either $\partial_r\Q=0$ or $f''(\Q) =0$. As justified in detail in Ref. \cite{DAmbrosio:2022}, both cases lead to no solutions beyond general relativity. Thus, solution set~1 of the connection does not lead to any new non-trivial solutions. Then, we proceed to solution set~2 of the connection.

\subsubsection{Off-diagonal metric equation for solution set 2}\label{ssec:SymRedFieldEq2}

We now turn to solution set 2, for which the off-diagonal metric equation is given by
\begin{equation}\label{eq:OffDiagSolset2}
    \M_{tr} = \frac12 \left(k +2 c\,(k-2c)\Gamma^{t}{}_{\theta\theta}\right) \partial_r\Q\, f''(\Q) = 0.
\end{equation}
From the discussion in the previous subsection, the equation is satisfied if
\begin{equation}\label{eq:AWayOut}
    k + 2 c\,(k-2c)\Gamma^{t}{}_{\theta\theta} = 0.
\end{equation}
The problem is therefore reduced to finding the solution of the above equation. This equation admits two classes of solutions:
\begin{itemize}
    \item \textbf{Class I:} $c = 0$ and $k = 0$.
    No connection component is explicitly determined.
    \item \textbf{Class II:}
    \begin{equation}\label{eq:Class2}
        \Gamma^{t}{}_{\theta\theta} =- \frac{k}{2c(k-2c)}\quad\text{for } c\neq 0\text{ and } k\neq 2c,
    \end{equation}
    such that $c$ and $k$ are constant parameters, and the component $\Gamma^{t}{}_{\theta\theta} $ is determined as a constant parameter. Then, using the first differential equation in solution set 2 (Table~\ref{tab:SolutionSet2}), another connection component is obtained as
    \begin{equation}
        \Gamma^{t}{}_{rr} = -\frac{k(8c^2+2ck-k^2)}{8c^2(k-2c)^2(\Gamma^{r}{}_{\theta\theta})^2}.
    \end{equation}
    The expression is well defined due to the condition $\Gamma^{r}{}_{\theta\theta}(r) \neq 0$ in solution set~2.
\end{itemize}
In summary, the off-diagonal metric equation in solution set~2 leads to two distinct classes of solutions, presented in Table~\ref{tab:Class}. It is noted that the results of Ref. \cite{DAmbrosio:2022} can be recovered when the component $\Gamma^{t}{}_{\theta\theta}$ in Class I is taken to be constant.

\begin{table}[h!]
    \centering
    \begin{tabular}{|p{0.16\textwidth}|p{0.29\textwidth} |p{0.28\textwidth} p{0.22\textwidth}|}
        \hline
        & & & \\[-1.5ex]
        & \textbf{Class I}& \textbf{Class II}&  \\[1ex]
        \hline
        & & & \\[-1.5ex]
        \textbf{Non-zero components}& $\Gamma^{t}{}_{\phi\phi} = \sin^2\theta\, \Gamma^{t}{}_{\theta\theta}$& $\Gamma^{t}{}_{tt} = (k-2c)/2$&$\Gamma^{t}{}_{tr} = \frac{k(k-4c)}{4c(k-2c)\Gamma^{r}{}_{\theta\theta}}$  \\[1ex]
        & $\Gamma^{r}{}_{\phi\phi} = \sin^2\theta\, \Gamma^{r}{}_{\theta\theta}$& $\Gamma^{t}{}_{rr} = -\frac{k(8c^2+2ck-k^2)}{8c^2(k-2c)^2(\Gamma^{r}{}_{\theta\theta})^2}$& $\Gamma^{t}{}_{\theta\theta} = -\frac{k}{2c(k-2c)}$\\[1ex]
        &  $\Gamma^{\theta}{}_{r\theta} = -1/\Gamma^{r}{}_{\theta\theta}$ & $\Gamma^{t}{}_{\phi\phi} = -\frac{k\sin^2\theta}{2c(k-2c)}$& $\Gamma^{r}{}_{tt} = c(k-2c)\Gamma^{r}{}_{\theta\theta}$\\[1ex]
        & $\Gamma^{\theta}{}_{\phi\phi} = -\cos\theta\,\sin\theta$& $\Gamma^{r}{}_{tr} = c+k/2$& $\Gamma^{r}{}_{\phi\phi} = \sin^2\theta\, \Gamma^{r}{}_{\theta\theta}$\\[1ex]
        &$\Gamma^{\phi}{}_{r\phi} = -1/\Gamma^{r}{}_{\theta\theta}$ &$\Gamma^{\theta}{}_{t\theta} = c$ & $\Gamma^{\theta}{}_{r\theta} = -\frac{k-4c}{2(k-2c)\Gamma^{r}{}_{\theta\theta}}$\\[1ex]
        &$\Gamma^{\phi}{}_{\theta\phi} = \cot\theta$ & $\Gamma^{\theta}{}_{\phi\phi} = -\cos\theta\,\sin\theta$& $\Gamma^{\phi}{}_{t\phi} = c$\\[1ex]
        &  For $\Gamma^{r}{}_{\theta\theta}(r)\neq 0$.& $\Gamma^{\phi}{}_{r\phi} = -\frac{k-4c}{2(k-2c)\Gamma^{r}{}_{\theta\theta}}$ & $\Gamma^{\phi}{}_{\theta\phi} = \cot\theta$ \\ [1ex]
        & & \multicolumn{2}{l|}{Arbitrary real constants $c\neq 0$ and $k\neq 2c$, together } \\[1ex]
        & & \multicolumn{2}{l|}{with the condition $\Gamma^{r}{}_{\theta\theta}(r)\neq 0$.} \\
        & & & \\[-1.5ex]
        \hline
        & & & \\[-1.5ex]
        \textbf{Differential relations}& $\partial_r \Gamma^t{}_{\theta\theta} = -\frac{\Gamma^t{}_{\theta\theta}+\Gamma^t{}_{rr}(\Gamma^r{}_{\theta\theta})^2}{\Gamma^r{}_{\theta\theta}}$&  \multicolumn{2}{l|}{$\partial_r \Gamma^r{}_{\theta\theta} = \frac{8c^2+k^2}{4c(k-2c)}-\Gamma^r{}_{rr}\Gamma^r{}_{\theta\theta}.$}\\[1ex]
        &$\partial_r \Gamma^r{}_{\theta\theta} = -1 -\Gamma^r{}_{rr}\Gamma^r{}_{\theta\theta}.$ & & \\[1ex]
        \hline
    \end{tabular}
    \caption{Classification of the connection components into Class I and Class II for solution set 2. }
    \label{tab:Class}
\end{table}

Let us now consider the temporal component of the connection field equations for solution set 2, which can be written as follows:
\begin{equation}
    \mathcal C_t = A\left(k+2c(k-2c)\Gamma^{t}{}_{\theta\theta}\right) - B\,c(k-2c)\partial_r\Gamma^{t}{}_{\theta\theta} = 0,
\end{equation}
where $A$ and $B$ are functions of the metric components, their derivatives, the derivatives of the non-metricity scalar and  $f(\Q)$. It is evident that the temporal component of the connection is also satisfied in both Class I and Class II. Thus, at this stage two of the field equations for both the metric $\M_{tr}$ and the connection $\C_{t}$ are solved. In what follows, we present the strategy for solving the remaining components of the field equations. These include the metric equations $\M_{tt}$, $\M_{rr}$, $\M_{\theta\theta}$, and the connection equation $\C_{r}$.

\subsubsection{Consistency of the equations for Class I and II}\label{sssec:Consistency}

Here, inspired by Ref. \cite{DAmbrosio:2022}, we solve the remaining field equations based on each class in Table~\ref{tab:Class}. In doing this, the aim is to provide new vacuum solutions beyond general relativity. The remaining components of the field equations are $\M_{tt}$, $\M_{rr}$, $\M_{\theta\theta}$ and $\C_{r}$, which shall be solved consistently with the differential equations listed in Table~\ref{tab:Class}. The component $\Gamma^r{}_{rr}$ appears in the differential equations in Table~\ref{tab:Class} and also in some of the remaining field equations. In both classes, this component is related to the non-metricity scalar by the following expressions:
\begin{align}\label{eq:Q}
	\mathbf{ClassI:}\quad \Q=& \frac{2}{r^{2}g_{rr}\Gamma ^{r}{}_{\theta \theta }%
	}\left( \frac{r^{2}}{\Gamma ^{r}{}_{\theta \theta }}+r(2+r\Gamma
	^{r}{}_{rr})+\Gamma ^{r}{}_{\theta \theta }-g_{rr}\Gamma ^{r}{}_{rr}\left(
	\Gamma ^{r}{}_{\theta \theta }\right) ^{2}-(r^{2}-g_{rr}\left( \Gamma
	^{r}{}_{\theta \theta }\right) ^{2})\right.   \notag   \\
	& \left. \times \frac{\partial _{r}g_{rr}}{2g_{rr}}+(r^{2}+2r\Gamma
	^{r}{}_{\theta \theta }+g_{rr}\left( \Gamma ^{r}{}_{\theta \theta }\right)
	^{2})\frac{\partial _{r}g_{tt}}{2g_{tt}}\right) ,  \notag \\
	\mathbf{ClassII:}\quad \Q=& \frac{2}{r^{2}g_{rr}}-\frac{2\Gamma
		^{r}{}_{rr}\Gamma ^{r}{}_{\theta \theta }}{r^{2}}-\frac{k(k+4c)}{%
		2cr^{2}(k-2c)}-\frac{r(k^{2}+8c^{2})+4c\Gamma ^{r}{}_{\theta \theta
		}(k-c)(2-r\Gamma ^{r}{}_{rr})}{4rg_{tt}}  \notag \\
	& +\frac{(k-4c)^{2}(r\left[ k^{2}+8c^{2}\right] -4c\Gamma ^{r}{}_{\theta
			\theta }\left[ k-2c\right] \left[ 2+r\Gamma ^{r}{}_{rr}\right] )}{%
		16rc^{2}(k-2c)^{2}g_{rr}\left( \Gamma ^{r}{}_{\theta \theta }\right) ^{2}}%
	+\left( \frac{(k-4c)^{2}}{c(k-2c)\Gamma ^{r}{}_{\theta \theta }}\right.  
	\notag \\
	& \left. +4g_{rr}\Gamma ^{r}{}_{\theta \theta }\left[ \frac{2}{r^{2}}-\frac{%
		c(k-2c)}{g_{tt}}\right] \right) \frac{\partial _{r}g_{rr}}{8(g_{rr})^{2}}%
	+\left( \frac{2}{r}-\frac{(k-4c)^{2}}{8c(k-2c)\Gamma ^{r}{}_{\theta \theta }}%
	+\frac{g_{rr}\Gamma ^{r}{}_{\theta \theta }}{2}\right.   \notag \\
	& +\left. \times \left[ \frac{2}{r^{2}}+\frac{c(k-2c)}{g_{tt}}\right]
	\right) \frac{\partial _{r}g_{tt}}{g_{tt}g_{rr}}.
\end{align}
Hence, instead of using the component $\Gamma^r{}_{rr}$ in the differential equations, one may employ $\Q$. In the following, it will be seen that $\Q$ is governed by a differential equation.

Now, our strategy to obtain the final result of the full set of equations is as follows
\begin{enumerate}
    \item By choosing one of the classes in Table~\ref{tab:Class}, the simplified form of all the field equations is obtained.
    \item According to \eqref{eq:Q}, the component $\Gamma^r{}_{rr}$ is directly extracted from $\Q$.
    \item By substituting $\Gamma^r{}_{rr}$ obtained in step (2) into all the field equations and the differential relations listed in Table~\ref{tab:Class}, $\Q$ enters as an independent quantity.
    \item For both classes, the metric field equations $\M_{tt}$ and $\M_{rr}$ contain first-order derivatives of the metric components, which have the following functional form:
    \begin{align}\label{eq:DE1}
        \partial_r g_{tt} &= \text{function}_1(g_{tt}, g_{rr},\Q, \partial_r\Q, \Gamma ^{r}{}_{\theta \theta })\notag\\
        \partial_r g_{rr} &= \text{function}_2(g_{tt}, g_{rr},\Q, \partial_r\Q, \Gamma ^{r}{}_{\theta \theta }).
    \end{align}
    \item By employing the first-order derivatives of the metric derived in Step~4, the differential equation for $\Gamma ^{r}{}_{\theta \theta }$ in both classes listed in Table~\ref{tab:Class} can be expressed as
    \begin{equation}\label{eq:DE2}
        \partial_r \Gamma ^{r}{}_{\theta \theta } = \text{function}_3(g_{tt}, g_{rr},\Q, \Gamma ^{r}{}_{\theta \theta }).
    \end{equation}
    Here, we do not take into account the first differential relation in Class I, due to the absence of $\Gamma^t{}_{\theta\theta}$ in the metric field equations.
    \item The metric field equation $\M_{\theta\theta}$ involves the first-order derivatives of the metric components and the second derivative $\partial _{r}^{2}g_{tt}$. Based on Step 4 and by differentiating $\partial_r g_{tt}$, the second derivative of $\Q$ can be written as:
    \begin{align}\label{eq:DE3}
        \partial^2_r g_{tt} &= \text{function}_4(g_{tt}, g_{rr}, \Q, \partial_r \Q, \Gamma ^{r}{}_{\theta \theta }) &\Longleftrightarrow & & \partial^2_r \Q &= \text{function}_5(g_{tt}, g_{rr}, \Q, \partial_r \Q, \Gamma ^{r}{}_{\theta \theta }).
    \end{align}
    This form is the same for both classes of solutions.
    \item Based on the results from Steps 4-6, the connection equation $\C_{r}$ is trivially satisfied in both classes.
\end{enumerate}
Through this strategy, one can solve all field equations for the solution sets in Class I and Class II. The four coupled differential equations govern $g_{tt}$, $g_{rr}$, $\Gamma ^{r}{}_{\theta \theta }$ and $\Q$. Once $f(\Q)$ is specified, this system can be solved to obtain non-trivial solutions. In the following section, we  calculate the solutions corresponding to two different choices of $f(\Q)$. We first obtain the solutions that reproduce general relativity and then extend them by including higher-order correction terms.

\section{Exact and approximate analytical solutions}\setcounter{equation}{0}\label{sec:Solutions}

In this section, we aim to solve all the field equations by obtaining the solutions of Eqs.~\eqref{eq:DE1}-\eqref{eq:DE3} for both Class I and Class II presented in Table~\ref{tab:Class}. We first find the vacuum solutions that are equivalent to those of general relativity. To do so, we set the non-metricity scalar $\Q$ to zero and solve the resulting field equations \cite{DAmbrosio:2022}. Then, we derive more general approximate analytical solutions beyond General Relativity by including a higher-order term of the non-metricity scalar in the Lagrangian. Specifically, it will be shown that the Class~I solutions reduce under certain conditions to those introduced in Ref. \cite{DAmbrosio:2022}.

\subsection{Class I}\label{ssec:ClassI}

Now, based on the strategy outlined in the previous section, the remaining field equations for the Class I solution set in Table~\ref{tab:Class} take the following form
\begin{align}\label{eq:metricClass1}
    \partial _{r}g_{tt}& =g_{tt}\left( -\frac{2-g_{rr}(2-\Q\,r^{2})}{2r}+\frac{%
        rg_{rr}f(\Q)\,}{2f^{\prime }(\Q)}+\frac{r^{2}-g_{rr}(\Gamma ^{r}{}_{\theta
            \theta })^{2}}{r\Gamma ^{r}{}_{\theta \theta }\,\,f^{\prime }(\Q)}(\partial
    _{r}\Q)\,f^{\prime \prime }(\Q)\right) ,  \notag \\
    \partial _{r}g_{rr}& =g_{rr}\left( \frac{2-g_{rr}(2-\Q\,r^{2})}{2\,r}-\frac{%
        r^{2}g_{rr}f(\Q)\,}{2f^{\prime }(\Q)}+\frac{r^{2}+2\,r\,\Gamma
        ^{r}{}_{\theta \theta }+g_{rr}(\Gamma ^{r}{}_{\theta \theta })^{2}\,}{%
        r\,\Gamma ^{r}{}_{\theta \theta }\,f^{\prime }(\Q)}(\partial _{r}\Q%
    )\,f^{\prime \prime }(\Q)\right) ,  \notag \\
    \partial _{r}\Gamma ^{r}{}_{\theta \theta }& =\frac{\Gamma ^{r}{}_{\theta
            \theta }}{2\left( r^{2}-g_{rr}(\Gamma ^{r}{}_{\theta \theta })^{2}\right) }%
    \left( 2r+g_{rr}(2-\Q\,r^{2})\left( r+2\Gamma ^{r}{}_{\theta \theta }\right)
    +r^{2}g_{rr}\left( r+\Gamma ^{r}{}_{\theta \theta }\right) \frac{f(\Q)}{%
        \,f^{\prime }(\Q)}\right) ,
\end{align}
together with
\begin{align}\label{eq:QClass1}
    \partial _{r}^{2}\Q=& \frac{g_{rr}(\partial _{r}\Q)}{2(r^{2}-g_{rr}\left(
        \Gamma ^{r}{}_{\theta \theta })^{2}\,\right) ^{2}}\biggl(-r^{3}(2-\Q%
    \,r^{2})+(\Gamma ^{r}{}_{\theta \theta })^{2}\left[ 6r+g_{rr}(2-\Q%
    \,r^{2})(3r+4\Gamma ^{r}{}_{\theta \theta })\right]   \notag \\
    & \left. -\left( r^{3}-g_{rr}(\Gamma ^{r}{}_{\theta \theta })^{2}(3r+2\Gamma
    ^{r}{}_{\theta \theta })\right) \frac{r^{2}f(\Q)}{f^{\prime }(\Q)}-\left(
    r^{2}-\,g_{rr}(\Gamma ^{r}{}_{\theta \theta })^{2}\right) \left(
    2r^{2}+g_{rr}\Gamma ^{r}{}_{\theta \theta }(2r+\Gamma ^{r}{}_{\theta \theta
    })\right) \right.   \notag \\
    & \times \frac{f^{\prime \prime }(\Q)}{g_{rr}f^{\prime }(\Q)}(\partial _{r}\Q%
    )-2(r^{2}-g_{rr}\left( \Gamma ^{r}{}_{\theta \theta })^{2}\,\right) ^{2}%
    \frac{f^{(3)}(\Q)}{g_{rr}f^{\prime \prime }(\Q)}(\partial _{r}\Q)\biggl).
\end{align}
Initially, by setting the non-metricity scalar $\Q$ to zero, the differential equation \eqref{eq:QClass1} is trivially satisfied, and hence we have
\begin{align}
    \partial _{r}g_{tt}& =-\frac{g_{tt}}{2r}\left( 2-g_{rr}\left( 2+\frac{%
        f(0)r^{2}}{\,f^{\prime }(0)}\right) \right),   \notag \\
    \partial _{r}g_{rr}& = \frac{g_{rr}}{2r}\left( 2-g_{rr}\left( 2+%
    \frac{f(0)r^{2}}{\,f^{\prime }(0)}\right) \right),   \notag \\
    \partial _{r}\Gamma ^{r}{}_{\theta \theta }& =\frac{\Gamma
        ^{r}{}_{\theta \theta }}{2\left( r^{2}-g_{rr}(\Gamma ^{r}{}_{\theta \theta
        })^{2}\right) }\left( 2r+2g_{rr}\left( r+2\Gamma ^{r}{}_{\theta \theta
    }\right) +\left( r+\Gamma ^{r}{}_{\theta \theta }\right) r^{2}g_{rr}\frac{%
        f(0)}{\,f^{\prime }(0)}\right).
\end{align}
Here, no specific form of $f$ has been chosen yet. Hence, the solutions to these equations read
\begin{align}
    g_{tt}& =1-\frac{2M}{r}+\frac{\,\Lambda _{\mathsf{eff}}}{3}r^{2},\qquad
    \qquad g_{rr}=\frac{1}{\,g_{tt}},  \notag \\
    \Gamma ^{r}{}_{\theta \theta }& =-r-\frac{\,\Lambda _{\mathsf{eff}}}{6}r^{2}+%
    \frac{c_{1}}{12f^{\prime }(0)}\pm \sqrt{\left( \frac{c_{1}}{12f^{\prime }(0)}%
        \right) ^{2}-r(6+\Lambda _{\mathsf{eff}}r^{2})\frac{c_{1}}{3f^{\prime }(0)}%
        +4r(72M-\Lambda _{\mathsf{eff}}r^{5})},
\end{align}
where $\Lambda _{\mathsf{eff}}=f(0)/\left( 2f^{\prime }(0)\right) $ is an effective cosmological constant and $c_{1}$ is an integration constant. This resembles the familiar Schwarzschild-de Sitter solution, and by fixing the constant parameter as $c_{1} = 12M f'(0)$, it leads to the following expression
\begin{align}
    g_{tt}& =1-\frac{2M}{r}+\frac{\,\Lambda _{\mathsf{eff}}}{3}r^{2},\qquad
    \qquad g_{rr}=\frac{1}{\,g_{tt}},  \notag \\
    \Gamma ^{r}{}_{\theta \theta }& =
    \begin{cases}
        -r \\[2mm]
        -r+2M-\Lambda _{\mathsf{eff}}r^{3}/3.
    \end{cases}
\end{align}
The solution with $\Gamma^{r}{}_{\theta\theta} = -r$ is identical to those previously given in Refs. \cite{Zhao:2022,Lin:2021,DAmbrosio:2022}. Thus, we find two more general families of solutions, which can reproduce the result of general relativity. In the following, we first consider $f(\Q) = \Q$, for which the corresponding solution is shown below. Then, we obtain approximate analytical asymptotically flat solutions for the extended case $f(\Q) = \Q + \alpha\,\Q^2$.

Regarding the case $f(\Q) = \Q$, the solutions to the field equations are of the form
\begin{align}
    g_{tt}& =1-\frac{2M}{r},\qquad \qquad g_{rr}=\frac{1}{\,g_{tt}},  \notag \\
    \Gamma ^{r}{}_{\theta \theta }& =-r+\frac{c_{1}}{12}\pm \frac{1}{12}\sqrt{%
        24r(12M-c_{1})+c_{1}^{2}},
\end{align}
which correspond to the standard Schwarzschild metric. Looking beyond this standard case, we include a higher-order correction term of the form $f(\Q) = \Q + \alpha\,\Q^2$ with $\alpha$ being a small parameter, i.e. $\left\vert \alpha \right\vert \ll 1$. To find the metric solutions for this form of gravity from Eqs.~\eqref{eq:metricClass1} and \eqref{eq:QClass1}, we expand the functions in powers of the small parameter $\alpha$ as follows
\begin{align}\label{eq:perturbClass1}
    g_{tt}& =g_{tt}^{(0)}+\alpha \,g_{tt}^{(1)}+\alpha ^{2}\,g_{tt}^{(2)},  \notag
    \\
    g_{rr}& =g_{rr}^{(0)}+\alpha \,g_{rr}^{(1)}+\alpha ^{2}\,g_{rr}^{(2)},  \notag
    \\
    \Gamma ^{r}{}_{\theta \theta }& =\gamma ^{(0)}+\alpha \,\gamma ^{(1)}+\alpha
    ^{2}\,\gamma ^{(2)},  \notag \\
    \Q& =\alpha \,\Q^{(1)}+\alpha ^{2}\,\Q^{(2)},
\end{align}
where the zeroth order terms are given by
\begin{align}
    g_{tt_\pm}^{(0)}& =1-\frac{2M}{r},\qquad \qquad g_{rr_\pm}^{(0)}=\frac{1}{\,g_{tt_\pm}^{(0)}},
    \notag \\
    \gamma ^{(0)}_\pm& =-r+\frac{c_{1}}{12}\pm \frac{1}{12}\sqrt{%
        24r(12M-c_{1})+c_{1}^{2}}.
\end{align}
Note that the two branches of the solution are denoted by the subscripts ``$+$'' and ``$-$''. Upon inserting \eqref{eq:perturbClass1} into the field equations, we proceed to solve the resulting equations order by order in $\alpha$. At zeroth order in $\alpha$, the field equations are trivially satisfied. At first order, the metric field equations decouple from the others and read
\begin{align}
    \partial_r g_{tt_+}^{(1)} = \frac{2 M\,r\, g_{tt_+}^{(1)} +(r-2M)^2\, g_{rr_+}^{(1)}}{(r-2M)\, r^2},&\qquad \qquad \partial_r g_{tt_-}^{(1)} = \frac{2 M\,r\, g_{tt_-}^{(1)} +(r-2M)^2\, g_{rr_-}^{(1)}}{(r-2M)\, r^2}, \notag \\
    \partial_r g_{rr_\pm}^{(1)} &= \frac{(2M+r)\, g_{rr_\pm}^{(1)}}{(2 M - r)\,r},
\end{align}
where their solutions are given by
\begin{equation}\label{eq:FirstOrderSolsClass1}
    g_{tt_\pm}^{(1)} = \frac{-c_2 + c_3\,(r-2M)}{r},\qquad \qquad g_{rr_\pm}^{(1)} = \frac{c_2 \, r}{(r-2M)^2}.
\end{equation}
Here, $c_2$ and $c_3$ are real constant parameters. In order to obtain asymptotically flat solutions, we fix $c_3=0$. Thus, the metric functions up to first order in $\alpha$ can be written as
\begin{equation}
    g_{tt_\pm}=1-\frac{2M_\text{ren}}{r},\qquad \qquad
    g_{rr_\pm}=\frac{1}{g_{tt_\pm}},
\end{equation}
where $2M_\text{ren} = 2M+\alpha\, c_2$ represents the renormalized Schwarzschild mass. Then, we move on to the next order of the metric field equations.

At second order in $\alpha$, using $g_{tt}^{(1)}$ and $g_{rr}^{(1)}$ for the two branches, the metric field equations are given by
\begin{align}\label{eq:secondOEMetP}
    \partial _{r}g_{tt_{+}}^{(2)}& =-\frac{c_{2}^{2}}{r(r-2M)^{2}}+\frac{%
        2Mg_{tt_{+}}^{(2)}}{r(r-2M)}+\frac{\left( r-2M\right) g_{rr_{+}}^{(2)}}{r^{2}%
    }-\frac{\sqrt{24r(12M-c_{1})+c_{1}^{2}}}{3r}\partial _{r}\,\Q^{(1)}_+  \notag
    \\
    \partial _{r}g_{rr_{+}}^{(2)}& =-\frac{c_{2}^{2}r}{(r-2M)^{4}}-\frac{\left(
        r+2M\right) g_{rr_{+}}^{(2)}}{r(r-2M)}-\frac{r\left( 24M-c_{1}\right) }{%
        3(r-2M)^{2}}\partial _{r}\,\Q^{(1)}_+,
\end{align}
and
\begin{align}\label{eq:secondOEMetM}
    \partial _{r}g_{tt_{-}}^{(2)}& =-\frac{c_{2}^{2}}{r(r-2M)^{2}}+\frac{%
        2Mg_{tt_{-}}^{(2)}}{r(r-2M)}+\frac{\left( r-2M\right) g_{rr_{-}}^{(2)}}{r^{2}%
    }+\frac{\sqrt{24r(12M-c_{1})+c_{1}^{2}}}{3r}\partial _{r}\,\Q^{(1)}_-  \notag
    \\
    \partial _{r}g_{rr_{-}}^{(2)}& =-\frac{c_{2}^{2}r}{(r-2M)^{4}}-\frac{\left(
        r+2M\right) g_{rr_{-}}^{(2)}}{r(r-2M)}-\frac{r\left( 24M-c_{1}\right) }{%
        3(r-2M)^{2}}\partial _{r}\,\Q^{(1)}_-.
\end{align}
To solve the above equations, we first need to determine $\partial _{r}\,\Q^{(1)}_\pm$. The differential equations governing $\partial_{r}\Q^{(1)}_{\pm}$ are derived as follows
\begin{equation}
    \frac{\partial _{r}^{2}\,\Q_{\pm}^{(1)}}{\partial _{r}\,\Q_{\pm}^{(1)}}=-\frac{1%
    }{r}+\frac{c_{1}^{2}}{2r\left( 24r(12M-c_{1})+c_{1}^{2}\right) }-\frac{r
        \sqrt{24r(12M-c_{1})+c_{1}^{2}}\pm 2\left( Mc_{1}+r\left[ 12M-c_{1}\right]
        \right) }{2r(r-2M)\sqrt{24r(12M-c_{1})+c_{1}^{2}}}.
\end{equation}
By solving the above, we can get
\begin{equation}
    \partial _{r}\,\Q^{(1)}_\pm =\frac{144Mc_{4}}{24r(12M-c_{1})+c_{1}^{2}\mp\left(
        12r-c_{1}\right) \sqrt{24r(12M-c_{1})+c_{1}^{2}}},
\end{equation}
where $c_{4}$ is real constant parameter. Having $\partial_{r}\Q^{(1)}_{\pm}$ at hand, we can solve Eqs.\eqref{eq:secondOEMetP} and \eqref{eq:secondOEMetM}, and hence obtain
\begin{align}\label{eq:GSecondOrderP}
    g_{tt_{+}}^{(2)} =&\frac{-6c_{5}+c_{6}\left( r-2M\right) -2c_{4}\sqrt{%
            24r(12M-c_{1})+c_{1}^{2}}-2c_{4}\left( 24M-c_{1}\right) \left( 1+2\ln
        (c_{1})\right) }{6r}  \notag \\
    & +\frac{2c_{4}\left( r(12M-c_{1})+Mc_{1}\right) }{6rM}\ln \left(\frac{%
        6r(12M-c_{1})+6M\left( c_{1}+\sqrt{24r(12M-c_{1})+c_{1}^{2}}\right) }{r-2M}\right),
    \notag \\
    g_{rr_{+}}^{(2)}=& \frac{r\left( c_{2}^{2}+c_{5}(r-2M)\right) }{(r-2M)^{3}}-%
    \frac{c_{4}r(24M-c_{1})}{3(r-2M)^{2}}\ln \left(\frac{6r(12M-c_{1})+6M\left( c_{1}+%
        \sqrt{24r(12M-c_{1})+c_{1}^{2}}\right) }{c_{1}^{2}(r-2M)}\right),
\end{align}
and
\begin{align}\label{eq:GSecondOrderN}
    g_{tt_{-}}^{(2)} =&\frac{-6c_{5}+c_{6}\left( r-2M\right) +2c_{4}\sqrt{%
            24r(12M-c_{1})+c_{1}^{2}}-2c_{4}\left( 24M-c_{1}\right) \left( 1-\ln
        (c_{1}^{2}/3)\right) }{6r}  \notag \\
    & -\frac{2c_{4}\left( r(12M-c_{1})+Mc_{1}\right) }{6rM}\ln \left(2(12M-c_{1})+%
    \frac{2M(c_{1}+\sqrt{24r(12M-c_{1})+c_{1}^{2}})}{r}\right),  \notag \\
    g_{rr_{-}}^{(2)} =&\frac{r\left( c_{2}^{2}+c_{5}(r-2M)\right) }{(r-2M)^{3}}+%
    \frac{c_{4}r(24M-c_{1})}{3(r-2M)^{2}}\ln \left(\frac{6r(12M-c_{1})+6M(c_{1}+\sqrt{%
            24r(12M-c_{1})+c_{1}^{2}})}{c_{1}^{2}r}\right).
\end{align}
Also, $c_{5}$ and $c_{6}$ are integration constants. Next, we examine the asymptotic behavior of the solutions in the cases of $c_{1} =12M$ and $c_{1} \neq12M$.

\begin{itemize}
    \item[$\blacksquare$] \begin{center}
        $\boldsymbol{c_{1} =12M}$
    \end{center}
\end{itemize}

For the case $c_{1} = 12M$, the second-order contribution to the metric takes the form
\begin{align}
    g_{tt_{+}}^{(2)}& =\frac{-6c_{5}+c_{6}\left( r-2M\right)
        -48c_{4}M-24c_{4}M\ln (r-2M)}{6r},  \notag \\
    g_{rr_{+}}^{(2)}& =\frac{r[ c_{2}^{2}+c_{5}(r-2M)-4c_{4}M(r-2M)\ln
        (r-2M)] }{(r-2M)^{3}},
\end{align}
and
\begin{align}\label{eq:SecondmetricN}
    g_{tt_{-}}^{(2)}& =\frac{-6c_{5}+c_{6}\left( r-2M\right) +24c_{4}M\ln (r)}{6r
    },  \notag \\
    g_{rr_-}^{(2)}& =\frac{r[ c_{2}^{2}+c_{5}(r-2M)-4c_{4}M(r-2M)\ln
        (r)] }{(r-2M)^{3}}.
\end{align}
The asymptotic flatness condition imposes $c_{6} = 0$ for both cases. As a result, the full metric components with the subscript ``$+$'' can be written as follows
\begin{equation}
    g_{tt_{+}}=1-\frac{2M_{ren_{+}}}{r}-\alpha ^{2}\frac{\mu }{r}\ln \left(
    \frac{r-2M}{r^{\ast }}\right),\qquad \qquad g_{rr_{+}} =\frac{1}{g_{tt_+}}+\frac{2\alpha ^{2}\mu r}{(r-2M)^{2}},
\end{equation}
where $2M_{ren_{+}}=2M+\alpha c_{2}+\alpha ^{2}\left( 8Mc_{4}+c_{5}\right)$ is the renormalized Schwarzschild mass and $\mu =4Mc_{4}$. The scale $r^{\ast
}$ arises from a shift in $c_{5}\longrightarrow c_{5}-4Mc_{4}\ln \left( r^{\ast
}\right)$ to ensure a dimensionless argument inside the logarithmic term.

Expanding the metric for large $r$ leads to
\begin{equation}\label{eq:eqhorizP}
    g_{tt_{+}}=1-\frac{2M_{ren_{+}}}{r}-\alpha ^{2}\frac{\mu }{r}\ln \left(
    \frac{r}{r^{\ast }}\right) +\alpha ^{2}\frac{2M\mu }{r^{2}}+%
    \mathcal{O}\left( \frac{1}{r^{3}}\right) ,
\end{equation}
in which the logarithmic term gives the correction beyond general relativity. On the other hand, one can determine the location of the horizon from the roots of $g_{tt_{+}} = 0$. Hence, we have
\begin{align}
    g_{tt_{+}} =0\,\, &\Longrightarrow \,\,   r-2M_{ren_{+}}-\alpha ^{2}\mu \ln \left( \frac{r-2M}{%
        r^{\ast }}\right)=0.
\end{align}
The solution of the above equation involves the Lambert function $\mathcal{W}$, which is provided in Appendix \ref{app:A}. Depending on the sign of $\mu$, the equation has two real or one roots. Two roots exist if
\begin{equation}
    0<\frac{r^{\ast }}{\alpha ^{2}\mu }e^{-\frac{c_{2}+\alpha \left(
            8Mc_{4}+c_{5}\right) }{\alpha \mu }}\leq e^{-1},
\end{equation}
in which the roots are written as
\begin{align}
    r_{Ih}&=2M-\alpha ^{2}\mu \mathcal{W}\left( -\frac{r^{\ast }}{\alpha ^{2}\mu }%
    e^{-\frac{c_{2}+\alpha \left( 8Mc_{4}+c_{5}\right) }{\alpha \mu }}\right)
    \approx 2M+r^{\ast }e^{-\frac{c_{2}+\alpha \left( 8Mc_{4}+c_{5}\right) }{%
            \alpha \mu }}   \notag \\
    r_{Oh}&=2M-\alpha ^{2}\mu \mathcal{W}\left( -1,-\frac{r^{\ast }}{\alpha
        ^{2}\mu }e^{-\frac{c_{2}+\alpha \left( 8Mc_{4}+c_{5}\right) }{\alpha \mu }%
    }\right) \approx 2M+\alpha c_{2}+\alpha ^{2}\left( 8Mc_{4}+c_{5}+\mu \ln
    \left( \frac{\alpha c_{2}}{r^{\ast }}\right) \right).
\end{align}
Here, $r_{Ih}$ and $r_{Oh}$ denote the radii of the inner and outer horizons, respectively. The approximations are achieved for small values of the parameter $\alpha$, indicating deviations from the Schwarzschild horizon radius. In the case of negative $\mu$ ( $\mu<0$), the equation \eqref{eq:eqhorizP} admits only one real root,
\begin{equation}
    r_{h}=2M+\alpha ^{2}\mu \mathcal{W}\left( \frac{%
        r^{\ast }}{\alpha ^{2}\mu }e^{\frac{c_{2}+\alpha \left( 8Mc_{4}+c_{5}\right)
        }{\alpha \mu }}\right) \approx 2M+\alpha c_{2}+\alpha ^{2}\left(
    8Mc_{4}+c_{5}+\mu \ln \left( \frac{\alpha c_{2}}{r^{\ast }}\right) \right),
\end{equation}
which also showing the deviation from the Schwarzschild horizon for small $\alpha$.

For the solution with the subscript ``$-$'' \eqref{eq:SecondmetricN}, the full form of the metric components is expressed as follows
\begin{equation}
    g_{tt_-} =1-\frac{2M_{ren_{-}}}{r}+\alpha ^{2}\frac{\mu }{r}\ln \left( \frac{%
        r}{r^{\ast }}\right), \qquad \qquad g_{rr_-} =\frac{1}{g_{tt_-}},
\end{equation}
where $2M_{ren_{-}}=2M+\alpha c_{2}+\alpha ^{2}c_{5}$ denotes the renormalized Schwarzschild mass, while $\mu =4Mc_{4}$. The scale $r^{\ast}$ arises through the redefinition $c_{5}\longrightarrow c_{5}+4Mc_{4}\ln \left( r^{\ast
}\right) $ to have a dimensionless argument inside the logarithm. This result is exactly the same as that found in Ref. \cite{DAmbrosio:2022}. The horizon radius is determined from $g_{tt_{-}} = 0$, yielding
\begin{equation}
    r-2M_{ren_{-}}+\alpha ^{2}\mu \ln \left( \frac{r}{r^{\ast }}%
    \right) =0,
\end{equation}
whose solution involves the Lambert $\mathcal{W}$ function. According to Appendix \ref{app:A}, for $\mu > 0$, the equation yields a single real root
\begin{equation}
    r_{h}=\alpha ^{2}\mu \mathcal{W}\left( \frac{r^{\ast
    }}{\alpha ^{2}\mu }e^{\frac{2M+\alpha c_{2}+\alpha ^{2}c_{5}}{\alpha ^{2}\mu }}\right) \approx
    2M+\alpha c_{2}+\alpha ^{2}\left( c_{5}-\mu \ln \left( \frac{2M}{r^{\ast }}
    \right) \right),
\end{equation}
where the deviation from the Schwarzschild horizon radius is clearly seen. While for $\mu < 0$, when
\begin{equation}
    -e^{-1}\leq \frac{r^{\ast }}{\alpha ^{2}\mu }e^{\frac{2M+\alpha c_{2}+\alpha
            ^{2}c_{5}}{\alpha ^{2}\mu }}<0,
\end{equation}
the two roots are given by
\begin{align}
    r_{Ih}&=\alpha ^{2}\mu \mathcal{W}\left( \frac{r^{\ast }}{\alpha ^{2}\mu }e^{%
        \frac{2M+\alpha c_{2}+\alpha ^{2}c_{5}}{\alpha ^{2}\mu }}\right) \approx r^{\ast }e^{\frac{2M+\alpha c_{2}+\alpha ^{2}c_{5}}{\alpha ^{2}\mu }},    \notag \\
    r_{Oh}&=\alpha ^{2}\mu \mathcal{W}\left( -1,\frac{r^{\ast }}{\alpha ^{2}\mu }%
    e^{\frac{2M+\alpha c_{2}+\alpha ^{2}c_{5}}{\alpha ^{2}\mu }}\right) \approx 2M+\alpha
    c_{2}+\alpha ^{2}\left( c_{5}-\mu \ln \left( \frac{2M}{r^{\ast }}\right)
    \right) .
\end{align}
The two inner and outer horizons are located at $r_{Ih}$ and $r_{Oh}$, respectively. Such results do not appear in Ref. \cite{DAmbrosio:2022}. In the following, we present the asymptotically flat solutions Eqs. \eqref{eq:GSecondOrderP} and \eqref{eq:GSecondOrderN} for the case $c_{1} \neq 12M$.

\begin{itemize}
    \item[$\blacksquare$] \begin{center}
        $\boldsymbol{c_{1} \neq12M}$
    \end{center}
\end{itemize}

Here, we examine the asymptotically flat behavior of the solutions Eqs. \eqref{eq:GSecondOrderP} and \eqref{eq:GSecondOrderN} for $c_{1} \neq 12M$. For the metric solution with the subscript ``$+$'', the asymptotic flatness condition requires that
\begin{equation}
    c_{6} =-\frac{2c_{4}(12M-c_{1})}{M}\ln \left( 6(12M-c_{1}) \right),
\end{equation}
so that the full metric components up to second order in $\alpha$ are given by
\begin{align}
	g_{tt_{+}}=& 1-\frac{2M}{r}-\frac{\alpha c_{2}}{r}-\alpha ^{2}\frac{%
		3c_{5}+c_{4}\sqrt{24r(12M-c_{1})+c_{1}^{2}}}{3r}-\frac{\alpha
		^{2}c_{4}\left( 24M-c_{1}\right) }{3r}  \notag \\
	& \times \left( 1-\ln \left( \frac{6r^{\ast }\left( 12M-c_{1}\right) }{%
		c_{1}^{2}}\right) \right) +\frac{\alpha ^{2}c_{4}\left(
		r(12M-c_{1})+Mc_{1}\right) }{3rM}  \notag \\
	& \times \ln \left( \frac{r(12M-c_{1})+M(c_{1}+\sqrt{24r(12M-c_{1})+c_{1}^{2}%
		})}{(r-2M)(12M-c_{1})}\right) , \\
	g_{rr_{+}}=& \frac{r}{r-2M}+\frac{\alpha c_{2}r}{(r-2M)^{2}}+\frac{\alpha
		^{2}r\left( c_{2}^{2}+c_{5}(r-2M)\right) }{(r-2M)^{3}}-\frac{\alpha
		^{2}c_{4}r(24M-c_{1})}{3(r-2M)^{2}}  \notag \\
	& \times \ln \left( \frac{6rr^{\ast }(12M-c_{1})+6Mr^{\ast }(c_{1}+\sqrt{%
			24r(12M-c_{1})+c_{1}^{2}})}{c_{1}^{2}(r-2M)}\right) .
\end{align}
In the above, the scale $r^{\ast}$ is introduced by shifting $c_{5}\longrightarrow c_{5}-c_{4}\left( 24M-c_{1}\right) \ln \left( r^{\ast
}\right) /3$ to make the argument of the logarithm dimensionless. Asymptotically, $g_{tt_{+}}$ behaves as follows
\begin{equation}
    g_{tt_{+}}=1-\frac{2M_{ren_{+}}}{r}+\alpha ^{2}\frac{\mu }{r^{3/2}}+
    \mathcal{O}\left( \frac{1}{r^{2}}\right),
\end{equation}
where
\begin{align}
    2M_{ren_{+}}=&2M+\alpha c_{2}+\alpha ^{2}\left( 4Mc_{4}+c_{5}\right) -\alpha
    ^{2}\frac{c_{4}(24M-c_{1})}{3}\ln \left( \frac{6r^{\ast }(12M-c_{1})}{%
        c_{1}^{2}}\right), \notag
\\
\mu =&\frac{16\sqrt{2}M^{2}c_{4}}{\sqrt{%
		3(12M-c_{1})}}.
	\end{align}
The location of the horizon is determined as
\begin{equation}
    g_{tt_{+}}=0\Longrightarrow r_{h}=2M+\alpha c_{2}+\alpha ^{2}\left( c_{5}+%
    \frac{c_{4}\left( 24M-c_{1}\right) }{3}\left[ 2+\ln \left( \frac{\alpha
        c_{2}c_{1}^{2}}{6r^{\ast }M\left( 24M-c_{1}\right) }\right) \right] \right) .
\end{equation}
This reveals the deviation from the Schwarzschild horizon radius up to second order in $\alpha$.

Similarly, we now analyze the behavior of the metric solution with the subscript ``$-$'' given in \eqref{eq:GSecondOrderN} for the case $c_{1} \neq 12M$. The asymptotic flatness condition leads to the following result
\begin{equation}
    c_{6}=\frac{2c_{4}(12M-c_{1})}{M}\ln \left( 2(12M-c_{1})\right).
\end{equation}
Hence, the metric components up to second order in $\alpha$ take the form
\begin{align}
	g_{tt_{-}}=& 1-\frac{2M}{r}-\frac{\alpha c_{2}}{r}-\alpha ^{2}\frac{%
		3c_{5}-c_{4}\sqrt{24r(12M-c_{1})+c_{1}^{2}}}{3r}-\frac{\alpha
		^{2}c_{4}\left( 24M-c_{1}\right) }{3r}  \notag \\
	& \times \left( 1+\ln \left( \frac{6r^{\ast }\left( 12M-c_{1}\right) }{%
		c_{1}^{2}}\right) \right) -\frac{\alpha ^{2}c_{4}\left(
		r(12M-c_{1})+Mc_{1}\right) }{3rM}  \notag \\
	& \times \ln \left( 1+\frac{M(c_{1}+\sqrt{24r(12M-c_{1})+c_{1}^{2}})}{%
		r(12M-c_{1})}\right) , \\
	g_{rr_{-}}=& \frac{r}{r-2M}+\frac{\alpha c_{2}r}{(r-2M)^{2}}+\alpha ^{2}%
	\frac{r\left( c_{2}^{2}+c_{5}(r-2M)\right) }{(r-2M)^{3}}+\alpha ^{2}\frac{%
		c_{4}r(24M-c_{1})}{3(r-2M)^{2}}  \notag \\
	& \times \ln \left( \frac{6rr^{\ast }(12M-c_{1})+6Mr^{\ast }(c_{1}+\sqrt{%
			24r(12M-c_{1})+c_{1}^{2}})}{c_{1}^{2}r}\right) .
\end{align}
The scale $r^{\ast}$ naturally arises through the redefinition $c_{5}\longrightarrow c_{5}+c_{4}\left( 24M-c_{1}\right) \ln \left( r^{\ast}\right) /3$. For $r \to \infty$, we have the asymptotic behavior
\begin{equation}
    g_{tt_{-}}=1-\frac{2M_{ren_{-}}}{r}-\alpha ^{2}\frac{\mu }{r^{3/2}}+\mathcal{%
        O}\left( \frac{1}{r^{2}}\right) ,
\end{equation}
where the parameters are defined as
\begin{align}
    2M_{ren_{-}} =&2M+\alpha c_{2}+\alpha ^{2}\left( 4Mc_{4}+c_{5}\right)
    +\alpha ^{2}\frac{c_{4}(24M-c_{1})}{3}\ln \left( \frac{6r^{\ast }(12M-c_{1})%
    }{c_{1}^{2}}\right), \notag \\
    \mu  =&\frac{16\sqrt{2}M^{2}c_{4}}{\sqrt{3(12M-c_{1})}}.
\end{align}
Finally, the location of the horizon is given by the root of the function $g_{tt_{-}}$, which can be expressed up to second order in $\alpha$ as
\begin{equation}
    r_{h}=2M+\alpha c_{2}+\alpha ^{2}\left( c_{5}+\frac{c_{4}\left(
        24M-c_{1}\right) }{3}\ln \left( \frac{6r^{\ast }\left( 24M-c_{1}\right) }{%
        c_{1}^{2}}\right) \right) .
\end{equation}
In the next part, we attempt to find solutions to all the field equations for Class II listed in Table~\ref{tab:Class} and explore the possible existence of asymptotically flat metric solutions.

\subsection{Class II}\label{ssec:ClassII}

Now, using the Class~II solution set given in Table~\ref{tab:Class}, we obtain the corresponding remaining field equations from Eqs.~\eqref{eq:DE1}-\eqref{eq:DE3}. The resulting metric field equations can be written in the following form
\begin{align}\label{eq:MetricComClass2}
    \partial _{r}g_{tt} =&-\frac{g_{tt}\left( 2+g_{rr}(\Q\,r^{2}-2)\right) }{2r}+%
    \frac{g_{tt}g_{rr}f(\Q)\,r}{2f^{\prime }(\Q)}-\frac{rg_{tt}(\partial _{r}\Q%
        )\,f^{\prime \prime }(\Q)}{\Gamma ^{r}{}_{\theta \theta }\,\,f^{\prime }(\Q)}%
    \left( \frac{(k-4c)^{2}}{8c(k-2c)}+\frac{g_{rr}(\Gamma ^{r}{}_{\theta \theta
        })^{2}\,}{r^{2}}\right.   \notag \\
    & \left. -\frac{c(k-2c)g_{rr}(\Gamma ^{r}{}_{\theta \theta })^{2}\,}{2g_{tt}}%
    \right),   \notag \\
    \partial _{r}g_{rr} =&\phantom{+}\frac{g_{rr}\left( 2+g_{rr}(\Q%
        \,r^{2}-2)\right) }{2\,r}-\frac{g_{rr}^{2}f(\Q)\,r^{2}}{2f^{\prime }(\Q)}+%
    \frac{rg_{rr}(\partial _{r}\Q)\,f^{\prime \prime }(\Q)}{8g_{tt}\,f^{\prime }(%
        \Q)}\left[ g_{tt}\,\left( \frac{16}{r}-\frac{(k-4c)^{2}}{c(k-2c)\Gamma
        ^{r}{}_{\theta \theta }}\right) \right.   \notag \\
    & \left. +4g_{rr}\Gamma ^{r}{}_{\theta \theta }\left( c(k-2c)+\frac{2g_{tt}}{%
        r^{2}}\right) \right] ,
\end{align}
while the other two equations read
\begin{align}\label{eq:ConnectionNonClass2}
	\partial _{r}\Gamma ^{r}{}_{\theta \theta }=& \frac{g_{rr}\Gamma
		^{r}{}_{\theta \theta }}{2\left( (k-4c)^{2}r^{2}g_{tt}-4c(k-2c)g_{rr}(\Gamma
		^{r}{}_{\theta \theta })^{2}(c(k-2c)r^{2}-2g_{tt})\right) }\left[
	(k-4c)^{2}\right.   \notag   \\
	\times & (2+g_{rr}(2-\Q\,r^{2}))\frac{g_{tt}}{\,g_{rr}}-16c(k-2c)g_{tt}(2-\Q%
	\,r^{2})\Gamma ^{r}{}_{\theta \theta }+4c^{2}(k-2c)^{2}r  \notag \\
	& \times (6-g_{rr}(2-\Q\,r^{2}))(\Gamma ^{r}{}_{\theta \theta })^{2}+\frac{%
		r^{2}f(\Q)}{\,f^{\prime }(\Q)}(g_{tt}((k-4c)^{2}r-8c(k-2c)\Gamma
	^{r}{}_{\theta \theta })  \notag \\
	& -4c^{2}(k-2c)^{2}rg_{rr}(\Gamma ^{r}{}_{\theta \theta })^{2})\biggr], 
	\notag \\
	\partial _{r}^{2}\Q=& \mathcal{A}\left( c,k,\Q,\partial _{r}\Q%
	,g_{tt},g_{rr},\Gamma ^{r}{}_{\theta \theta },f(\Q),f^{\prime }(\Q%
	),f^{\prime \prime }(\Q),f^{(3)}(\Q)\right) .
\end{align}
Due to the complicated expression of $\mathcal{A}$, its explicit form is not shown here. To avoid this complexity and difficulty, we first consider the case $f(\Q) = \Q$, which reproduces general relativity. It is worth stressing again that $c$ and $k$ are constant parameters, see \eqref{eq:Class2}. To obtain the Schwarzschild solution within Class~II, we set $\Q=0$ in which case the field equations reduce to the form
\begin{align}
    \partial _{r}g_{tt}& =-\frac{g_{tt}}{r}\left( 1-g_{rr}\right) ,\qquad \partial
    _{r}g_{rr}=\frac{g_{rr}}{r}\left( 1-g_{rr}\right),   \notag \\
    \partial _{r}\Gamma ^{r}{}_{\theta \theta }& =\frac{g_{tt}\Gamma
        ^{r}{}_{\theta \theta }\left( (k-4c)^{2}(1+g_{rr})r-16c(k-2c)g_{rr}\Gamma
        ^{r}{}_{\theta \theta }\right) +4c^{2}(k-2c)^{2}(3-g_{rr})rg_{rr}(\Gamma
        ^{r}{}_{\theta \theta })^{3}}{%
        (k-4c)^{2}r^{2}g_{tt}-4c(k-2c)(c(k-2c)r^{2}-2g_{tt})g_{rr}(\Gamma
        ^{r}{}_{\theta \theta })^{2}}.
\end{align}
Here, the equation governing $\Q$, the second of Eqs.~\eqref{eq:ConnectionNonClass2}, is automatically satisfied. Solving these equations yields the following expressions
\begin{align}
	g_{tt} =&1-\frac{2M}{r},\qquad \qquad \qquad g_{rr}=\frac{1}{\,g_{tt}}, 
	\notag \\
	\Gamma ^{r}{}_{\theta \theta } =&-\frac{\left( r-2M\right) }{2\left(
		r-2M\right) -c\left( k-2c\right) r^{3}}  \notag \\
	& \times \left( 2r-c_{1}\pm \sqrt{\left( 2r-c_{1}\right) ^{2}+\frac{r\left(
			k-4c\right) ^{2}\left( 2\left( r-2M\right) -c\left( k-2c\right) r^{3}\right) 
		}{4c\left( k-2c\right) }}\right) ,
\end{align}
with $M$ being the Schwarzschild mass and $c_{1}$ an integration constant. With the general relativity equivalent-solution at hand, we next explore its generalization in the symmetric teleparallel framework. In order to derive an analytical generalization, we restrict ourselves to the case $k = 4c$. By substituting $k = 4c$ into the above solution, we get
\begin{equation}
    g_{tt}=1-\frac{2M}{r},\qquad \qquad g_{rr}=\frac{1}{\,g_{tt}},  \qquad \qquad
    \Gamma ^{r}{}_{\theta \theta }=
    \begin{cases}
        0 \\
        -\left( r-2M\right) \left( 2r-c_{1}\right)/
        (r-2M-c^{2}r^{3}),
    \end{cases}
\end{equation}
where $\Gamma ^{r}{}_{\theta \theta } = 0$ is not allowed according to Table~\ref{tab:Class}. Similar to the Class~I generalization, the general relativity Lagrangian is extended by a quadratic term in $\Q$, namely $f(\Q) = \Q + \alpha\,\Q^2$, where $\lvert \alpha \rvert \ll 1$. To solve Eqs.~\eqref{eq:MetricComClass2} and \eqref{eq:ConnectionNonClass2}, we again assume the following series expansions in powers of $\alpha$
\begin{align}\label{eq:perturbClass2}
    g_{tt}& =g_{tt}^{(0)}+\alpha \,g_{tt}^{(1)}+\alpha ^{2}\,g_{tt}^{(2)},  \notag
    \\
    g_{rr}& =g_{rr}^{(0)}+\alpha \,g_{rr}^{(1)}+\alpha ^{2}\,g_{rr}^{(2)},  \notag
    \\
    \Gamma ^{r}{}_{\theta \theta }& =\gamma ^{(0)}+\alpha \,\gamma ^{(1)}+\alpha
    ^{2}\,\gamma ^{(2)},  \notag \\
    \Q& =\alpha \,\Q^{(1)}+\alpha ^{2}\,\Q^{(2)},
\end{align}
where at zeroth order, we have
\begin{equation}
    g_{tt}^{(0)}=1-\frac{2M}{r}\qquad \qquad g_{rr}^{(0)}=\frac{1}{\,g_{tt}^{(0)}%
    },\qquad \qquad\gamma ^{(0)}=-\frac{\left( r-2M\right) \left( 2r-c_{1}\right) }{%
        r-2M-c^{2}r^{3}}.
\end{equation}
By plugging the series \eqref{eq:perturbClass2} into the field equations \eqref{eq:MetricComClass2} and \eqref{eq:ConnectionNonClass2}, we solve the corresponding equations order by order in the $\alpha$ expansion. At zeroth order in $\alpha$, the equations are automatically fulfilled, whereas at first order, the metric field equations are given by
\begin{equation}
    \partial _{r}g_{tt}^{(1)}=\frac{2M\,r\,g_{tt}^{(1)}+(r-2M)^{2}\,g_{rr}^{(1)}%
    }{(r-2M)\,r^{2}},\qquad \qquad \partial _{r}g_{rr}^{(1)}=\frac{(2M+r)\,g_{rr}^{(1)}}{%
        (2M-r)\,r}.
\end{equation}
The above equations resemble those of Class~I, whose solutions are expressed as
\begin{equation}
    g_{tt}^{(1)}=\frac{-c_{2}+c_{3}\,(r-2M)}{r},\qquad \qquad g_{rr}^{(1)}=\frac{c_{2}\,r}{%
        (r-2M)^{2}},
\end{equation}
where $c_2$ and $c_3$ are real integration constants. Imposing asymptotic flatness implies $c_3=0$, and the metric components up to first order in $\alpha$ can be written as
\begin{equation}
    g_{tt}=1-\frac{2M_{\text{ren}}}{r},\qquad \qquad g_{rr}=\frac{1}{g_{tt}},
\end{equation}
where we have defined $ 2M_\text{ren} = 2M+\alpha\, c_2$. Proceeding to the second order, the metric field equations are found to be
\begin{align}
	\partial _{r}g_{tt}^{(2)} =&-\frac{c_{2}^{2}}{r(r-2M)^{2}}+\frac{%
		2Mg_{tt}^{(2)}}{r(r-2M)}+\frac{\left( r-2M\right) g_{rr}^{(2)}}{r^{2}}%
	+2\left( 2-\frac{c_{1}}{r}\right) \partial _{r}\,\Q^{(1)},  \notag \\
	\partial _{r}g_{rr}^{(2)} =&-\frac{c_{2}^{2}r}{(r-2M)^{4}}-\frac{\left(
		r+2M\right) g_{rr}^{(2)}}{r(r-2M)} \notag \\
	& +\frac{2r\left( 8M^{2}-2M\left( c_{1}+2r-2c^{2}r^{3}\right)
		+rc_{1}+c^{2}r^{3}\left( c_{1}-4r\right) \right) }{(r-2M)^{2}\left(
		r-2M-c^{2}r^{3}\right) }\partial _{r}\,\Q^{(1)},
\end{align}
where $\partial _{r}\,\Q^{(1)}$ obeys the following equation
\begin{equation}\label{eq:DnonmetClass2}
    \frac{\partial _{r}^{2}\,\Q^{(1)}}{\partial _{r}\,\Q^{(1)}}=-\frac{2\left(
        8M^{2}+r^{2}\left( 2-c^{2}rc_{1}\right) -Mr\left( 8-c^{2}r\left(
        3c_{1}-2r\right) \right) \right) }{\left( r-2M\right) \left( 2r-c_{1}\right)
        \left( r-2M-c^{2}r^{3}\right) }.
\end{equation}
The solution to this equation is
\begin{equation}
    \partial _{r}\,\Q^{(1)}=-\frac{c_{4}\left( r-2M-c^{2}r^{3}\right) }{\left(
        r-2M\right) \left( 2r-c_{1}\right) ^{2}},
\end{equation}
where $c_{4}$ is an integration constant. With this, the solution to Eq. \eqref{eq:DnonmetClass2} is found to be
\begin{align}
	g_{tt}^{(2)}=& c_{6}+2c^{2}c_{4}\left( r-2M\right) -\frac{c_{5}+2c_{6}M-c_{4}%
	}{r}+\frac{c_{4}c^{2}}{4r\left( 4M-c_{1}\right) }  \notag \\
	& \times \left( c_{1}^{2}\left( 12M+c_{1}\right) -16M^{2}\left(
	16M+c_{1}\right) \right) +\frac{16c_{4}c^{2}M^{2}}{\left( 4M-c_{1}\right)
		^{2}}\left( 10M-3c_{1}-\frac{M\left( 24M-7c_{1}\right) }{r}\right)   \notag
	\\
	 & \times \ln (r-2M)-\frac{c_{4}c^{2}c_{1}^{2}\left( 3M-c_{1}\right) \left(
		2r-c_{1}\right) }{r\left( 4M-c_{1}\right) ^{2}}\ln (2r-c_{1}), \notag \\
	g_{rr}^{(2)}=& \frac{rc_{2}^{2}}{(r-2M)^{3}}+\frac{r\left(
		2c_{5}+c^{2}r\left( 2r+4M+3c_{4}\right) \right) }{2(r-2M)^{2}}-\frac{%
		rc_{4}\left( 16M-4c_{1}+c^{2}c_{1}^{3}\right) }{4(r-2M)^{2}(2r-c_{1})} 
	\notag \\
	& +\frac{c_{4}c^{2}r}{(r-2M)^{2}(4M-c_{1})}\left( 16M^{3}\ln
	(r-2M)+c_{1}^{2}\left( 3M-c_{1}\right) \ln (2r-c_{1})\right) ,
\end{align}
in which $c_{5}$ and $c_{6}$ are integration constants. By expanding the metric components at infinity, an asymptotically flat case is achieved when $c_{4}=0$ and $c_{6}=0$. Thus, the asymptotically flat form of the metric components up to second order in $\alpha$ is given by
\begin{equation}
    g_{tt}=1-\frac{2M_{\text{ren}}}{r},\qquad \qquad g_{rr}=\frac{1}{g_{tt}},
\end{equation}
with $M_{\text{ren}}$ being the renormalized Schwarzschild mass, $2M_{\text{ren}}\ =2M+\alpha \,c_{2}+\alpha^{2}\,c_{5}$. Up to second order in $\alpha$, Class II does not admit any asymptotically flat solutions that go beyond general relativity.

\section{Conclusion and Outlook}\label{sec:Conclusion}
In this work, we have explored spherically symmetric vacuum
solutions in the context of symmetric teleparallel $f(\Q)$
gravity. Our starting point was a recent classification of affine
connections that respect both the symmetries of spacetime and the
core geometric assumptions of the theory-namely, the absence of
curvature and torsion. Considering the metric and the connection
as independent dynamical degrees of freedom, we derived the full
set of field equations and explored their consistency conditions
in ample detail.

One of our main goals in this work is to move beyond the general
relativistic limit and uncover solutions that carry the imprint of
non-metricity. To do so, we first identified two distinct classes
of connections that satisfy the off-diagonal metric equation and
the temporal component of the connection field equations. These
classes emerged naturally from the constraint analysis and
provided the foundation for all subsequent calculations. For both
classes, we began by considering the case where the non-metricity
scalar $\Q$ vanishes. In this limit, the field equations simplify
considerably, and we were able to recover exact solutions that are
fully equivalent to those of general relativity. These include the
Schwarzschild black hole as well as Schwarzschild (anti-)de Sitter
generalization, valid for any smooth function $f(\Q)$ satisfying
$f(0)\neq0$ and $f'(0)\neq0$. This interesting result confirms
that general relativity is smoothly embedded in the framework of
$f(\Q)$ gravity, and that the standard vacuum solutions remain
intact when non-metricity is switched off.

The more interesting regime, however, is the one in which
non-metricity plays an active role. To explore this, we focused on
the simplest nonlinear extension, namely $f(\Q)=\Q+\alpha ~\Q^2$,
with $\alpha$ treated as a small parameter. Working perturbatively
to second order in $\alpha$, we constructed asymptotically flat
solutions that extend the Schwarzschild metric in a nontrivial
way. These solutions introduce new integration constants that are
absent in general relativity. These constants can be interpreted
as hair associated with the affine connection. In this sense,
black holes in $f(\Q)$ gravity are not only characterized by their
mass, but also by the imprints of the non-metricity field. Let us
summarize two classes of solutions, separately.

For the first class of solutions, we obtained two distinct
branches of corrections, distinguished by the sign of a square
root term in the perturbative expansion. These branches exhibit
different asymptotic behaviors and, in some cases, give rise to
logarithmic corrections to the metric. Depending on the
parameters, the horizon radius can receive positive or negative
corrections relative to the Schwarzschild value, and in certain
regimes, inner and outer horizons can emerge. This means the
horizon structure becomes richer compared to general relativity.
Remarkably, the equation determining the horizon location reduces
to a transcendental form whose solution is naturally expressed in
terms of the Lambert $\mathcal{W}$ function. This result echoes
similar structures in other modified theories of gravity.

For the second class of solutions, the analysis was complicated by
the presence of additional constant parameters inherited from the
connection. Nevertheless, by restricting to a particular subclass
($k=4c$), we were able to obtain perturbative solutions that again
reveal corrections to the Schwarzschild geometry. These findings
suggest that the space of spherically symmetric solutions in
$f(\Q)$ gravity is considerably richer than previously
appreciated.

Throughout the analysis, we emphasized the importance of treating
the connection field equations on equal footing with the metric
equations. The connection is not merely an auxiliary structure,
but carries its own dynamics and can leave observable imprints on
the spacetime geometry. In our solutions, these imprints manifest
as deviations in the asymptotic behavior, the horizon radius, and
the overall functional form of the metric.

Finally, we would like to emphasis that the results presented in
this work open up several avenues worth exploring. On the
theoretical side, it would be natural to extend our perturbative
analysis to higher orders in $\alpha$ or to consider other
functional forms of $f(\Q)$ such as power-law, exponential, or
models with a cosmological constant. The aim is to explore whether
the connection hair we have identified persists or gives way to
new features. The stability of these solutions under linear
perturbations is another important question. On the observational
side, while deviations from Schwarzschild are small, their
logarithmic and Lambert function structure might leave imprints on
phenomena like black hole shadows, quasinormal modes, or the
gravitational wave ringdown. It would be interesting to explore
whether these effects could be constrained by current or future
observations. We hope that the solutions and methods presented
here will serve as a useful foundation for further exploration in
the context of $f(\Q)$ theory of gravity.
\acknowledgments{We thank Shiraz University Research Council. This
work is based upon research funded by Iran National Science
Foundation (INSF) under project No. 4028419.}

\appendix
\section{Analytical solutions using the Lambert function $\mathcal{W}$}
\label{app:A}

\begin{figure}
	\centering
	\includegraphics[width=0.57\textwidth]{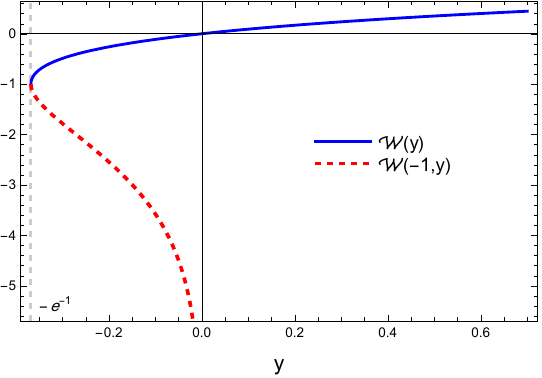}
	\caption{The real branches of the Lambert function, $\mathcal{W}(y)$ and $\mathcal{W}(-1,y)$. $\mathcal{W}(y)$ is indicated by the blue solid line, defined for $-e^{-1} \leq y < +\infty$. The red dashed line corresponds to $\mathcal{W}(-1,y)$, defined on the interval $-e^{-1} \leq y < 0$.}
	\label{fig:LambertFunction}
\end{figure}

The Lambert function $\mathcal{W}$ is defined as the solution of the general equation \cite{Corless:1996}
\begin{equation} \label{eq:LambertF}
    \mathcal{W}\left( y\right) e^{\mathcal{W}\left( y\right) }=y.
\end{equation}
The Lambert function admits two real branches, which are denoted here as $\mathcal{W}(y)$ and $\mathcal{W}(-1,y)$. The overall behavior of these two branches is illustrated in Figure~\ref{fig:LambertFunction}.
By multiplying Eq. \eqref{eq:LambertF} by the factor $a\,e^{-\mathcal{W}}$ with $a \neq 0$ and choosing $y=-e^{-b/a}/a$, we arrive at $a \mathcal{W}=-e^{-\mathcal{W}-b/a}$. Introducing the variable $\mathcal{W}=-\ln \left( x\right)-b/a$, one obtains
\begin{equation}
    x-a\ln \left( x\right) -b=0.
\end{equation}
Depending on the values of $a$ and $b$, the equation has one or two real solutions. When $0 < e^{-b/a}/a < e^{-1}$, the equation admits two real solutions, given by:
\begin{equation}
    x=
    \begin{cases}
        -a\mathcal{W}\left( -e^{-b/a}/a\right) \\[2mm]
        -a\mathcal{W}\left( -1,-e^{-b/a}/a\right),
    \end{cases}
\end{equation}
For $a \ll b$, the above solutions can be approximated as follows
\begin{equation}
    x\approx
    \begin{cases}
        e^{-b/a} \\[2mm]
        b+a\ln \left( b\right) +a^{2}/b\ln \left( b\right).
    \end{cases}
\end{equation}
If $a < 0$, there is a single solution of the following form
\begin{equation}
    x =-a\mathcal{W}\left(-e^{-b/a}/a\right).
\end{equation}
In the regime $\vert a \vert \ll b$, the solution behaves as
\begin{equation}
    x\approx b+a\ln \left( b\right) +\frac{a^{2}}{b}\ln \left( b\right).
\end{equation}


\end{document}